\documentstyle[12pt]{article}
\newcommand{\doublespacing}{\let\CS=\@currsize\renewcommand{\baselinesstrech}
{2.0}\tiny\CS}

\begin{document}
\setlength{\baselineskip}{18.5pt}
\thispagestyle{empty}\setcounter{page}{1}
                               
\centerline{\bf PLANCK SCALE PHYSICS , PREGEOMETRY AND THE NOTION }
\centerline{\bf  OF TIME} 

\vskip20pt
\noindent
\centerline{\bf S. ROY}
\centerline {Physics and Applied Mathematics Unit}
\centerline{Indian Statistical Institute, Calcutta- 700 108, India}
\centerline{e-mail : sisir@isical.ac.in}
\begin{center}
{\bf Abstract}
\end{center}
Recent progress in quantum gravity and string theory has raised interest among
scientists to whether or not nature behaves discretely at the Planck scale.
There are two attitudes twoards this discretenes i.e. top-down and bottom-up
approach. We have followed up the bottom-up approach. Here we have tried to
describe how macroscopic space-time or its underlying mesoscopic substratum
emerges from a more fundamental concept. The very concept of space-time, causality may not be valid beyond Planck scale. We have introduced the concept of
generalised time within the framework of Sheaf Cohomology where the physical
time emrges around and above Planck scale. The possible physical amd metaphysical implications are discussed.

\vskip20pt
\noindent  
{\bf 1. Introduction}
\vskip5pt
\noindent
 Recent progress in quantum gravity $[1]$ and string theory $[2]$ has raised 
interest among scientists to whether or not nature behaves ``discretely''
at the Planck scale.
However, it is not clear what this metaphor means or how it should be 
implemented into
systematic study concerning physics and mathematics in the Planck regime.

There are basically two attitudes towards  this discreteness at Planck scale. 
One starts from continuum concept and then tries to detect or create modes of ``discrete 
behaviour'' on finer scales. We call this as ``top down approach'' $[3]$. On the other hand, 
one can try to describe how macroscopic space-time or its underlying mesoscopic substratum emerges from 
a more fundamental concept like fluctuating cellular network $[4]$ around  the Planck 
scale. We call this  as  bottom up $[5]$ approach. It is generally believed that no 
physical laws which are valid in continuum space-time will be valid beyond or around the Planck scale.
The very concept of space-time, causality may not be valid beyond Planck scale. 
Some scientists suspect  space or time should be considered as  emergent 
properties $[6]$. 
\vskip5pt
\noindent
In a recent paper $[5]$ we developed a kind of pre geometry around the
Planck scale where notion of "clock time" is proposed  below or near the  Planck 
scale.
This is not to be confused with the physical time. It is possible to understand this time
if one defines a category of generalized time $[7]$ using the idea of sheaf cohomology. This
generalized time may not be necessarily linear in nature.
There is  no before or after in this category. Physical time with its usual characteristics emerges from this 
generalized time under appropriate conditions. . Then one gets
linearity and ordering
(like before and after) in the sequence of time. This ordering of the sequence can be thought as an emergent property.

In this paper, we shall briefly describe the developments of quantum gravity and string
theory so as to understand the conceptual problems related to the discrete behaviours at the 
Planck 
scale in section 2. In section 3, we shall describe the pre geometry, cellular networks and
overlapping fuzzy lumps. Then we shall introduce the concept of generalized time and the
emergence of physical time in section 4. Finally, a few remarks concerning the possible 
physical and 
philosophical implications have been made in section 5.
\vskip10pt
\noindent
{\bf 2. New  Challenges in 20th Century Physics}
\vskip5pt
\noindent
The birth of quantum mechanics in early twentieth century shattered our idea whether 
it is
possible to describe our physical world with one physical law. The world on large
scale - the motion of terrestrial objects like planetary motion, stars and galaxies
down to our day to day world are explained with the help of Newtonian mechanics. But as 
we move to the smaller scales, regarding the behaviour of objects with small masses like
electrons, protons and photons, one needs to consider different physical theory as quantum
theory. So we have two different types of physical law that are valid at two different 
levels of the physical world. However, the popular belief among the physicists 
is that
if we really understand quantum theory, it is possible to use it to describe the large-scale phenomena  
i.e. what  we call classical world. In practice, however, we use either
classical physics or  quantum physics. Penrose $[8]$ curiously noticed  `` how the ancient
Greeks looked at the world... one set of laws ... applied to earth and different set.. 
in the heavens.''
\vskip5pt
\noindent
Let us now look at the scales which we deal with  the physical world. The time and length 
scales at the
bottom are known as Planck time and the  Planck length respectively. Planck time is 
$10^{-43}$ sec  which is considered the shortest time in our physical world i.e. shorter than the shortest 
lived particles, called resonance, which is about $10^{-23}$ sec.
Likewise, the  Planck length which is about $10^{-33}$cm, is considered  the shortest length i.e.
fundamental unit
of length. Now, when one needs to combine both Planck time and Planck length, it is necessary
to consider both quantum theory and general  relativity. Quantum theory is valid 
for small
length scales and general  relativity  for large length, and time, scales. As soon as both. quantum theory and general relativity 
are brought together,
one needs to consider both Planck length and Planck time. For example, if we need to 
describe the
physics of black holes or the universe at the the big bang, it is necessary to
consider both quantum theory and general  relativity. But, the attempt to combine
the quantum formalism with general  relativity leads to catastrophe rather than the
harmony in nature.
\vskip5pt
\noindent
Twentieth-century researche in quantum gravity $[1]$  changed our 
focus into the physical world. Before merging of quantum theory and general
relativity, one needs to consider as the microscopic nature of space-time i.e. as we go to
the smaller and smaller length scales, the nature of space-time should be studied  
with uncertainty principle . From the point of view  of classical
mechanics, the space remains flat at arbitrary small 
lengths.
Quantum mechanics changes this conclusion. Everything in this world even the gravitational
field is subject to quantum fluctuations inherent in the uncertainty principle.
The uncertainty principle indicates that the size of the fluctuation of gravitational field becomes 
larger as we focus our
attention to the smaller regions of space. These quantum fluctuations might be manifested 
as 
violent distortions of the surrounding space and hence the curvature related to the 
gravitational
field. John Wheeler  $[9]$ described it as quantum foam in which the notions of left 
and right, before and after lose their meaning. At this scale, we face the fundamental
incompatibility
between general  relativity and quantum theory. Essentially, the notion of smooth
geometry that is prerequisite for general relativity is not a valid
concept at small scales due to violent fluctuations of the quantum world at. Now if we consider
 the distances used in everyday life, the random or violent fluctuations concel each other
 and the concept of smooth geometry is valid.
.This is similar to a picture composed of dots. When viewed from afar, the
picture presents a smooth image. However, if we view it from a very small distance or at finer 
scales, we see  nothing but a collection of dots, each separate from the other. This pictire's discrete structure appears upon examines at small scales.
The fabric of space-time looks similar i.e. discrete at the smallest scale (Planck length) 
and smooth at large scales.
\vskip5pt
\noindent
The  various contradictory attempts to incorporate gravity into a quantum mechanical framework leads
to new a search for deeper understanding of the nature. Green et al $[10]$
presented a convincing evidence that superstring theory might provide this
understanding. Strings are considered as one-dimensional filament like objects,
vibrating to and fro. Accordingly, the elementary ingredients of the universe are not now  
point like objects but one-dimensional filaments. String theory proposes
that strings are ultramicroscopic ingredients making up the particles from
which atoms themselves are made. On average, strings are of the size of Planck
length. Proposing strings as fundamental ingredients of  the
universe has far reaching consequences. In particular, String theory appears to resolve the
conflict between general  relativity and quantum mechanics.
\vskip5pt
\noindent
Despite its elegance, string theory has not so far led to any new predictions 
concerning the properties of elementary particles. Moreover, for strictly 
technical reasons string theory requires nine or even ten spatial dimensions. 
The emergence of these
extra spatial dimensions is probably
one of the most difficult issues, unless one appeals solely to mathematical formalism. 
In fact, it is very difficult to get an intuitive non technical 
reason 
for the existence of these extra dimensions.
 Rutherford once commented that if you can't explain the result in 
simple
non technical term, you really don't understand it. Strings are considered 
as fabric of space-time
like a piece of material out of which the universe is tailored.. This may give
rise to
new possibility to understand the space-time at small scales. But what do we
really mean by the fabric of the Universe?
This question has been debated for hundreds of years. We can summarize it
briefly as follows :

\begin{itemize}

\item  Newton declared space and time as the eternal and immutable  ingredients
in the make up of the cosmos.

\item  Leibniz [11] dissented, claiming that space and time are merely 
bookkeeping
devices for summarizing relationship between objects and events in the
universe. The location of an object in space and in time has meaning only in 
comparison with another object.

\item  Mach [12] further developed Leibniz's  view, which is much 
closer to
the view of today's physicists.
\end{itemize}

After the proposal of string theory, the imminent question arises  whether or not 
we should view
ourselves as truly being embedded in something when we refer to our immersion
in the space-time fabric.  However, the string theory does not suggest an
answer to this question. Moreover, an ordinary piece of fabric is the end product
of raw material being carefully woven. In the raw state, before strings, there is
no notion like ``before''. Maybe, our language is not so developed to handle
it. Is this world really comprehensible?
\vskip10pt
\noindent
{\bf 3. Planck Scale and Pregeometry}
\vskip10pt
\noindent
It is evident from the above analysis that  physicists do not have a satisfactory model of
the physical world at the Planck scale.
In our search for such a model, our working philosophy is that the continuum 
concepts of ordinary physics and/or mathematics can be reconstructed 
from more primordial pre geometric (basically
discrete) concepts , which are prevalent  at the  Planck scale. Here, geometry emerges from
a purely relational picture {\it a` la }  Leibniz. In particular,
the underlying substratum of our physical world or, more specifically the
space-time (quantum) vacuum  can be viewed as  cellular network $[4 ]$. This discrete 
structure
consists of elementary nodes, $n_i$, which interact or exchange information
with each other via bonds, $b_{ik}$, playing the role of irreducible
elementary interactions. The possible internal structures of the nodes, modules
or bonds (interaction channels) are described  by a  discrete internal state space
carried by the nodes/bonds. The node set is assumed to be large but finite or
countable. The bond $b_{ik}$ is assumed to connect the nodes $n_i, n_k$. 
The internal 
state of the  nodes/bonds are denoted by $s_i, J_{ik}$ respectively. Our
philosophy is to generate complex behaviours out of simple models.
\vskip5pt
\noindent
Let us choose
$$s_i \in q.{\bf Z} ,  \  \   J_{ik} \in {-1,0,+1}$$

with $q$ an elementary quantum of information. In our approach, the bond 
states are dynamical degrees of freedom which, {\it a fortiori}, can be switched 
off or on. The  wiring, that is the pure geometry of the network is also 
an emergent, dynamical property and is not given in advance.  Consequently, 
the nodes and the bonds are not arranged  
in any regular way i.e. a lattice, and there is no fixed  near/far order. 
This implies that geometry will become to some extent a relational (Machian) 
concept and is not {\it a priori} element. In this model, the nodes and bonds 
are updated in discrete  clock time steps, $t = z.\tau, z \in {\bf Z}$  and 
$\tau$ being an elementary clock time interval.
This updating is given by some local dynamical law.  Here, local means that the 
node/bond states are changed at each clock time step according to a prescription
with input to all states of some neighbourhood (in some topology) of
the node/bond under discussion. Here,  $t$ is not to be 
confused with the physical time, which is also expected  to be an 
emergent coarse grained quantity. 
\vskip5pt
\noindent
Now we shall introduce the concept of generalized
time  and its relation to physical time. 
\vskip5pt
\noindent
There are many different aspects of our class of cellular networks. Our cellular netorks can be regarded as
complex dynamical systems, or  statistical / stochastic
frameworks, but in purely geometric sense, they are evolving graphs. It is
possible to describe the evolution and structure of large dynamical graphs.
Furthermore, the network  at any clock time encodes the complete near and the far-order structure of the
network at other clock times. It tells us the relative proximity of network subsets in terms of  possible physical aspects
such as strength of correlations or  statistical
distance. Stochastic aspects result from the underlying network
law, which induces, among other things, a certain amount of creation and 
annihilation
of bonds among the nodes. As a consequence
the size and shape of the
cliques or lump fluctuates in  course of network evolution. This derived
coarser network i.e. the clique graph or web of lumps, is defined by using 
meta-nodes to represent the cliques and  meta-bonds 
to represent overlap of cliques.
\vskip5pt
\noindent
It is to be noticed that while this new network may be regarded as being 
coarser
in some sense, in general it may nevertheless  consists of many more nodes and 
bonds than
the underlying primordial network. Usually there are many more maximal 
subsimplices than 
primordial nodes, as a given node will typically belong to quite a few
different subsimplices. This array of intersecting maximal subsimplices has
the natural structure of a simplical complex with smaller simplices as faces of
maximal ones i.e. the cliques. If we represent this simplical complex by a 
new (clique)
graph with only the maximal simplices occurring as meta-nodes, we loose, on the 
other
side, some information, as we do not keep track of situations where, say, three
lumps or cliques have a common overlap. It is possible to make some kind of 
ensemble
averages over fluctuating but individual cliques and identify them with fuzzy 
lumps. The underlying philosophy is that the space (set of points) is replaced
by a class of functions on this space. This is similar to the philosophy of
non-commutative geometry.
The underlying graph carries a natural distance function
$$d(n_i,n_k)$$
i.e., the minimal length of a path connecting the given nodes where the length
of a path is simply the number of bonds comprising it. Keeping the
labeled nodes or cliques fixed, the distance fluctuates in  clock 
time. The clique metric will fluctuate since the cliques change their shape and
size i.e. also their degree of overlap. Now when we switch from the above kind
of dynamical picture of a time dependent graph, the ensemble picture of fuzzy
cliques or lumps, our point of view changes to a static but probabilistic one in the
spirit of Menger $[13]$. Here, the structure of the space is no longer time
dependent while its largely hidden dynamics is now encoded in various probabilistic
notions.
\vskip10pt
\noindent
{\bf 4. Pregeometry and Notion of Time}
\vskip5pt
\noindent
In the above analysis we have discussed two approaches to the geometric structure
at Planck scale. In one approach, we have considered underlying network with
dynamical laws and the static picture of fuzzy cliques or lumps with probabilistic
notion in other approach. It raises the age old dilemma regarding the physical world and
the mathematical laws. There are people who prefer to think of mathematical concepts
merely as idealization of our physical world. Here, the mathematical world is considered
to be emerging from the world of physical objects. The other group prefers to
think of the physical world as emerging out of the "timeless" world of mathematics.
Penrose $[8]$ remarked  "one of the remarkable things about the behaviour of the
world is how it seems to be grounded in mathematics to a quite extraordinary
degree of accuracy...more we understand about the physical world, and the deeper we
probe into the laws of nature, ... the physical world almost evaporates and we
are left only with mathematics".
Our working philosophy to understand the physical world around Planck scale is that
at this stage of development as physical laws beyond Planck scale are yet to found,
we shall start with mathematical concepts and mathematical laws and observe how the
physical world emerges.
\vskip5pt
\noindent
The recent developments in category
theory and sheaf cohomology $[7]$ shed new light at understanding the notion of
time below Planck length.  Category theory was created by Eilenberg and 
MacLane$[14]$
in the forties. It provides a powerful and very general methods in algebraic
geometry and algebraic analysis. 
\vskip5pt
\noindent
Let us start with some fundamental concepts
in category theory.
A category consists of objects and morphisms. 
For each pair of objects $X$ and $Y$ of $\bf C$, we have the set  ${\rm Hom}_{\bf C}(X,Y)$ of morphisms from
X to Y. For morphisms ${\it f}$ from X to Y and ${\it g}$ from Y to Z, the composition ${\it g} \ o \ {\it  f}$ 
of ${\it g}$ and  ${\it f}$ is defined and the composition ${\it g} \ o \ {\it f}$ is a morphism from  X to Z, 
which satisfies the associate law :

$$ {\it h} \ o  {({\it g} \ o {\it f})} =  {({\it h} \ o{\it g})} \ o \ {\it f}$$
For each object X, there is a morphism, called the identity morphism $id_x$, from  
X to X itself, satisfying $id_x \ o \ {\it f}  = {\it f}$ and and ${\it g} \ o\ id_x = {\it g}$ \ for any 
morphism of ${\it g}$ from X to X.

For example : (a) The category of sets  consists of objects being sets and 
morphisms being set theoretic maps. (b) The category of abelian groups 
as its objects and group homomorphism as its morphism.

Functor is an important concept in category theory.
A Functor $\bf F$ from a category $\bf C$ to category $\bf C^{\prime}$
is defined as :
for each object $X$  in  $\bf C$ ,
$\bf F$  assigns an object  ${\bf F} \ X$ in $\bf C^{\prime}$, such that
for each morphism  
$$f : X \rightarrow  Y {\rm in} \  {\bf C}$$
$\bf F$ assigns a morphism 
$$ {\bf F} {\it  f} : {\bf F} X \rightarrow {\bf F} Y \in  \bf C^{\prime}$$
Then $\bf F$ must satisfy 
$${\bf F} id_x = id_{{\bf F} X}$$ 
 and
 $$F{\it (gof)} = {\bf F} go{\bf F} {\it f}$$

A contravariant functor is a covariant functor from the dual category 
${\bf C}^{\rm opp}$ of $\bf C$ to $\bf C^{\prime}$.
Given a category $\bf C$, a new category ${\bf C}^{\rm opp}$, called the dual category
 of $\bf C$, is obtained in the following manner.
 
 (i) \ \ The objects of the category ${\bf C}^{\rm opp}$ coincide 
 with the objects of the category $\bf C$.
 
 (ii)\ \ The set of morphisms ${\rm Hom}_{ \bf C^{\rm opp}}(XY)$ is identical with
 ${\rm Hom} _{\bf C^{\rm opp}}(YX)$

 (iii) \ \ The composition map ${\rm Hom} {\bf C}^{\rm opp}(XY) x {\rm Hom} \bf C^{\rm opp}(YZ) 
 \rightarrow {\bf C}^{\rm opp}(X,Z)$
 
So $${(\bf C)^{\rm opp}} = {\bf C}$$

 \vskip5pt
 \noindent

The concept of dual category enables  one to dualize each notion and each statement
with respect to a category ${\bf C}$ into a notion and a statement with respect
to the category ${\bf C}^{\rm opp}$.

An important example is a presheaf.
The concept of presheaf  and sheaf play significant role in category theory.
The sheaf (cohomology) has been employed as a bridge from local information to
global information.
Let $\bf T$ be a topological space i.e. Euclidean n-space $\bf R^n$.
A presheaf $\it F$ is an assignment;
for every open subset $V$ of $\bf T$,
 $$V \rightarrow {\it F} V $$
where $\it F V$ is an object in a category satisfying some contravariant functor
 axioms.
Now if a presheaf is given, one can ask whether it is possible to obtain a global
information from a collection of local data by ``pasting'' those local data.
The answer is ``yes'' if the presheaf further satisfies some axiom. This means that
if a presheaf $\it F$ is actually a sheaf, then not only the discrete information
data ${\bf F(V_i)}$ can be obtained for each covering, but also global
information can be obtained by gluing the local data. 
\vskip5pt
\noindent
Let us now define $\hat T$ as a category of presheaves on the category $\bf T$
associated with a topological space $\it T$ with values in a product category
$\prod_{\alpha \in \Gamma} \bf C_{\alpha}$. More precisely, $\hat T$ is the category of 
contravariant
functors from the category $\bf T$ associated with a topological space $\it T$ 
to a product category $\prod_{\alpha \in \Gamma} \bf C_{\alpha}$ 
 where $\Gamma$ is an index set. The category $\bf T$ is said to
be the generalized time space or generalized time category $[7]$ when the real 
line $\bf R$
is embeddable in $\it T$. Namely,
$$\hat T =  {\prod_{\alpha \in \Gamma} \bf C_{\alpha}}^{{\bf T} ^{\rm opp}}$$.

To be more explicit, for an object $V$ in $\it T$ i.e. an open set $V$ of
$\it T$, and for an object $P$ in $\hat T$, we have $ P(V) = (P_{\alpha}(V)), \\ \alpha \in \Gamma$ 
where each $P_{\alpha}(V)$ is an object of $\bf C_{\alpha}$.
Recall that by the definition an entity is a presheaf $P$ in $\hat T$ where
 ${\bf C_{\alpha}, \alpha \in \Gamma}$ represents the totality of physical 
 categories. 
 It should be noted that
$\hat T$ includes matter like elementary particles, atoms, molecules etc.
The index set $\Gamma$ may be divided into several parts. The first part of
$\Gamma$ is used for physical world categories. We will use integers as indices 
for physical categories:
$$ \bf C_j, \  \   j = 0,1,2,....\in \Gamma$$
where $\bf C_0$ is the generalized time category $\it T$ itself, $\bf C_1$ is
the micro world and $\bf C_2$ is the macro world.  $\bf C_1 \& \bf C_2$ 
are discrete categories with structures.
Now, the time what we experience or use in physical world, is a linear and
uni-dimensional space i.e. the real line $R$. This allows us to introduce
the notion of before and after i.e. past , present and future.
We make the following assumption:

There exists an embedding 
$$ {\bf i} : \bf R \rightarrow \bf T$$. Here, the presheaf $P$ restricted to the closed subset ${\bf i}(\bf R)$ of $\bf T$
is a presheaf $i^{-P}$ over $i(R)$. In this world, several worlds exist simultaneously
with respect to generalized time.

The cohomology of network of entities can be  constructed in the following manner.
One defines a complex $\Sigma$ in a category say $\bf C$ as a sequence of
entities with morphisms in chain from one object to the other, the composition
of two consecutive morphisms being a zero morphism. The sequence 
$$\frac{\small \gamma}{}\!\!\!\rightarrow  P(U) \frac{\small \delta}{} \!\!\! \rightarrow Q(U)
\frac{\small \phi}{} \!\!\! \rightarrow R(U)\rightarrow..... $$

is such that this forms a cochain complex, namely, any consecutive composition
of morphisms in the above sequence is trivial.
The cohomology at $Q(U^{\prime})$  
denoted by $H^{\star}(....\rightarrow Q(U)\rightarrow...)$ is defined as the subquotient.

Now, if there is one entity $Q$, the above sequence becomes,
$$....\rightarrow 0 \rightarrow Q(U)\rightarrow 0\rightarrow.....$$.
Then the cohomology at $Q(U)$ is $Q(U)$ itself. That is the sub object of $Q(U)$ which
has no influence on anyone is the whole $Q(U)$.  Consider the above sequence for
two entities as
$$......\rightarrow 0\rightarrow P(U)\frac{\small \delta}{} \!\!\! \rightarrow
Q(U)\rightarrow 0\rightarrow...$$.
Then the cohomology at $Q(U)$ is the quotient $\frac{Q(U)}{Im \delta_{ij}}$ i.e., the
cohomology at $Q(U)$ is the quotient obtained by regarding the influence or
information $Q(U)$ receives from $P(U)$ as the trivial part of $Q(U)$.
On the other hand, the cohomology at $P(U)$ is the subobject ${\rm Ker} \delta_{ij}$.
Ker denotes the kernel and Im be the Imaginary part.
 In this manner one
can construct the cohomology for sequence of many entities and the influence of 
influence will not be lost.
\vskip10pt
\noindent
{\bf 5. Implications}
\vskip5pt
\noindent
The above analysis shows that it is possible to construct a pre geometric
framework so as to describe the physics around Planck scale. At the deepest level i.e.
beyond Planck level, we proposed purely mathematical concepts like category of 
generalized
time. As soon as we impose certain restriction or condition, we get physical time
around or above Planck scale. This generalized time category is called ``noumenon''
of infinite time and imposing restriction we get ``phenomenon'' time or physical
time. This conditioning is intimately related to defining measures and also to
perception. By using different ways of conditioning we can have the notion of time
at different scales or level of the physical world namely in classical mechanics,
theory of relativity or in quantum domain. So the physical time seems to be
emergent property due to conditioning or limitations.
\vskip5pt
\noindent
There is a great deal of debates specially at philosophical level about the 
concept of emergence.
Butterfield and Isham$[6]$ made an extensive review on this topic.
In everyday language, emergence is considered to be a process in temporal sense. But
in our framework at the level beyond Planck scale. In generalized time category
 there is no concept of ``before'' or ``after''. So ``emergence'' should be considered in
non-temporal sense. Here, ``emergence'' can be thought of associated with "conditioning"
or ``imposing limitation in measurement''. The reality beyond Planck scale  should
be considered as ``Veiled reality'' of d'Espagnat $[15]$. He suggested that it is
necessary to speak of an independent reality which can't be described in the
sense of traditional realism. One can get statistical knowledge of it.
\vskip5pt
\noindent
In Perennial Indian Philosophy $[16]$ this kind of  ``veiling'' is described by the
term  ``Maya'' or ``Illusion''.  In Sanskrit language the generic meaning of the 
word  ``Maya''
is related to  ``Measure'' and hence the limitation. Again the word "emergence"
has also been described in a completely different way rather than western sense.
It is said that ``emergence'' is related to the concept ``expansion from within
without''. This is not an expansion from a small center or focus but without 
reference to size or limitation
or area, means the development of ``limitless subjectivity into as limitless
objectivity''. Accordingly, this expansion not being an increase in size -for
infinite extension (generalized time) admits  no enlargement - is just a 
change of condition i.e. from pre geometry to discrete Planck scale and then to
continuum space-time. Here, the expansion is traced back to its origin to
a kind of primordial (and periodic) vibration which is responsible for
the manifestation of the physical objects and physical world. 
\vskip5pt
\noindent
It needs further
elaboration to find the linkage between the fluctuations associated with
the creation or annihilation of nodes or creation or annihilation of black holes
as discussed in our above cellular network model. This fluctuation might be
responsible for the fuzzy character of the lumps or foamy space -time as
speculated by Wheeler $[9]$.  
\vskip10pt
\noindent
{\bf References}
\vskip5pt
\noindent
\begin{enumerate}
\item  Isham  C.J.(1995) {\it Structural Issues in Quantum Gravity}, gr-qc/9510063.
\item  Smolin L.(1998) {\it The future of spin networks} in  S.A Huggett,   L.J.Mason , 
 K.P.Tod, S.T. Tsou, N.M.J.Woodhouse (eds),{\it The Geometrical          
Universe}, Oxford University Press, Oxford.
\item Rovelli  C.(1999)
 {\it Strings, loops \& others, a critical survey of the
present approaches to quantum gravity} in N.Dadhich and J.V.Narlikar(eds),
{\it Gravitation and Relativity : At the turn
of the millennium}, Poona University Press.    .
\item  Requardt M.(1998) {\it  Cellular Networks as Models for Planck scale
Physics}, J.Phys.A.Math.Gen.{\bf 31}, 7997-8021, hep-th/9806135.
\item  Requardt M. \& Roy S.(2001){\it (Quantum) space-time as statistical geometry of
fuzzy lumps and the connection with Random metric spaces},  Class. and Quantum. Gravity,
{\bf 18}, 3039-3058.
\item  Butterfield J.B., Isham  C.J.(1999){\it On the emergence of time in quantum gravity}
gr-qc/9901024.
\item  Kato G.(2002){\it Sheaf theoretic foundations of ontology}, preprint, January.
\item  Penrose R.(1999) in R.Penrose, N.Cartwright, S.W.Hawking and M. Longair(eds),
{\it The Large, The small and The Human Mind},
 Cambridge University Press (U.K.) pp.12-15.
\item  Wheeler J. and Ford K.(1998) {\it Geons, Black Holes and Quantum Foam}, 
W.W.Norton \&  Company,NY,London.
\item Green M.B.,Schwarz J.H.,Witten E.(1987) {\it Superstring Theory},  Cambridge
University Press, Cambridge.
\item Leibniz G.W. (1973) {\it The Monadology and The Leibniz  Clark-Leibniz
correspondance in Leibniz, Philosophical Writings},in Morris and 
G.H.R.Parkinson(trsls), Dent, London.
\item Mach E.(1960){\it The Science of Mechanics - A  Critical and Historical
Account of the Development}, Open  Court, La salle (originally published in 1886.)
\item Menger K.(1942){\it Statistical Metrics}, Procd. Nat. Acad.S C.USA
{\bf 28},535.
\item  MacLane S.(2000){\it Categories for the Working Mathematician}, 2nd ed. Springer
Verlag, New York.
\item d'Espagnat B. (1995){\it Veiled Reality, An Analysis of Present-Day Quantum 
Mechanical \bf Concepts}, Addision-Wesley, Reading(Mass)
\item Blavatsky H.(1979){\it The Secret Doctrine} Vol.I The Theosophical Publishing
House, Adyar, Madras, India, 7th Ed(originally published in 1888).
\end{enumerate}

\end{document}